\documentclass[%
 aip,apl
 sd,%
 amsmath,amssymb,
 reprint,%
]{revtex4-1}

\usepackage{graphicx}% Include figure files
\usepackage{dcolumn}% Align table columns on decimal point
\usepackage{bm}% bold math
\begin{document}

\preprint{AIP/123-QED}

\title{Scalable GaSb/InAs tunnel FETs with non-uniform body thickness}% Force line breaks with \\
%\thanks{Footnote to title of article.}

\author{Jun Z. Huang}
\email{junhuang1021@gmail.com}
\author{Pengyu Long}%
% \email{Second.Author@institution.edu.}
\author{Michael Povolotskyi}%
\author{Gerhard Klimeck}%
\affiliation{
Network for Computational Nanotechnology and Birck Nanotechnology Center, Purdue University, West Lafayette, IN 47907 USA%Authors' institution and/or address%\\This line break forced with \textbackslash\textbackslash
}%
\author{Mark J. W. Rodwell}
% \homepage{http://www.Second.institution.edu/~Charlie.Author.}
\affiliation{%
Department of Electrical and Computer Engineering, University of California at Santa Barbara, Santa Barbara, CA 93106-9560 USA%Second institution and/or address%\\This line break forced% with \\
}%

\date{\today}% It is always \today, today,

\begin{abstract}
GaSb/InAs heterojunction tunnel field-effect transistors are strong candidates in building future low-power integrated circuits, as they could provide both steep subthreshold swing and large ON-state current ($I_{\rm{ON}}$). However, at short channel lengths they suffer from large tunneling leakage originating from the small band gap and small effective masses of the InAs channel. As proposed in this article, this problem can be significantly mitigated by reducing the channel thickness meanwhile retaining a thick source-channel tunnel junction, thus forming a design with a non-uniform body thickness. Because of the quantum confinement, the thin InAs channel offers a large band gap and large effective masses, reducing the ambipolar and source-to-drain tunneling leakage at OFF state. The thick GaSb/InAs tunnel junction, instead, offers a low tunnel barrier and small effective masses, allowing a large tunnel probability at ON state. In addition, the confinement induced band discontinuity enhances the tunnel electric field and creates a resonant state, further improving $I_{\rm{ON}}$. Atomistic quantum transport simulations show that ballistic $I_{\rm{ON}}=284$A/m is obtained at 15nm channel length, $I_{\rm{OFF}}=1\times10^{-3}$A/m, and $V_{\rm{DD}}=0.3$V. While with uniform body thickness, the largest achievable $I_{\rm{ON}}$ is only 25A/m. Simulations also indicate that this design is scalable to sub-10nm channel length.

\end{abstract}

%                             % Classification Scheme.
%                              %display desired
\maketitle

%\section{\label{sec:level1}First-level heading:\protect\\ The line
%break was forced \lowercase{via} \textbackslash\textbackslash}
Tunnel field-effect transistor (TFET), a promising replacement of classical metal-oxide-semiconductor field-effect transistor (MOSFET) for future low-power integrated circuits, has been intensively studied over a decade. The advantages of TFET come from its steep subthreshold swing (SS) that overcomes the 60mV/dec limit of a conventional MOSFET, allowing substantial supply voltage ($V_{\rm{DD}}$) scaling \cite{ionescu2011tunnel}. However, because of low tunnel probability the steep SS usually occurs at very low current level \cite{seabaugh2010low,lu2014tunnel}. This leads to insufficient ON-state current ($I_{\rm{ON}}$) and thus large switching delay ($CV_{\rm{DD}}/I_{\rm{ON}}$). Various approaches have been proposed to improve the low $I_{\rm{ON}}$. In particular, GaSb/InAs heterojunction based TFETs can considerably boost $I_{\rm{ON}}$ due to their broken/staggered-gap band alignment \cite{mohata2011demonstration,lind2015iii}.

However, as the channel length scales to sub-20nm as projected by International Technology Roadmap for Semiconductors (ITRS) for the next technology nodes \cite{itrs_website}, the GaSb/InAs n-type TFETs suffer from large ambipolar and source-to-drain tunneling leakage due to the small band gap and the small effective masses of the InAs channel. These leakage can be reduced by reducing the body thickness \cite{avci2013heterojunction}, because the band gap and the effective masses of the InAs channel increase as the body thickness decreases. Meanwhile, the large band gap and large effective masses also reduce the tunneling probability across the tunnel junction. The resonant TFET with a reversed InAs/GaSb heterojunction can have a steep SS at short gate length \cite{avci2013heterojunction} but the $I_{\rm{ON}}$ is limited by the narrow resonant transmission peak \cite{long2016design}. The channel heterojunction design with a large band-gap AlInAsSb alloy as the channel material has also been proposed \cite{long2016high,long2016extremely} to mitigate the short-channel effects. However, a good-quality dielectric on top of AlInAsSb has not been experimentally demonstrated yet. Therefore, InP, a material with a high-quality dielectric already demonstated, has been investigated as the alternative channel material \cite{long2016iprm}. It is found that the lattice-mismatched InP channel imposes biaxial compressive strain on the GaSb/InAs tunnel junction, compromising the improvement.

\begin{figure}
\includegraphics[width=4.25cm]{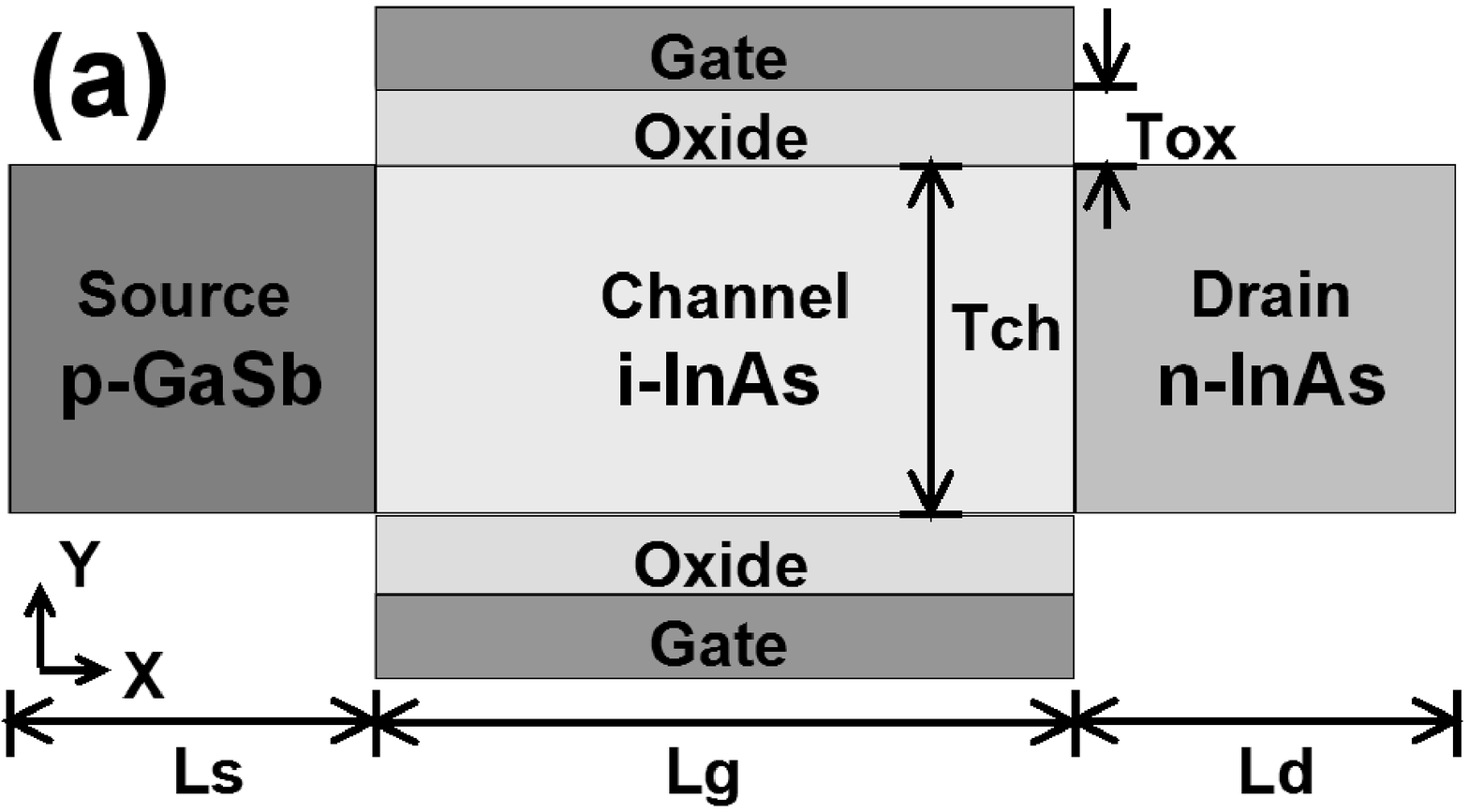}
\includegraphics[width=4.25cm]{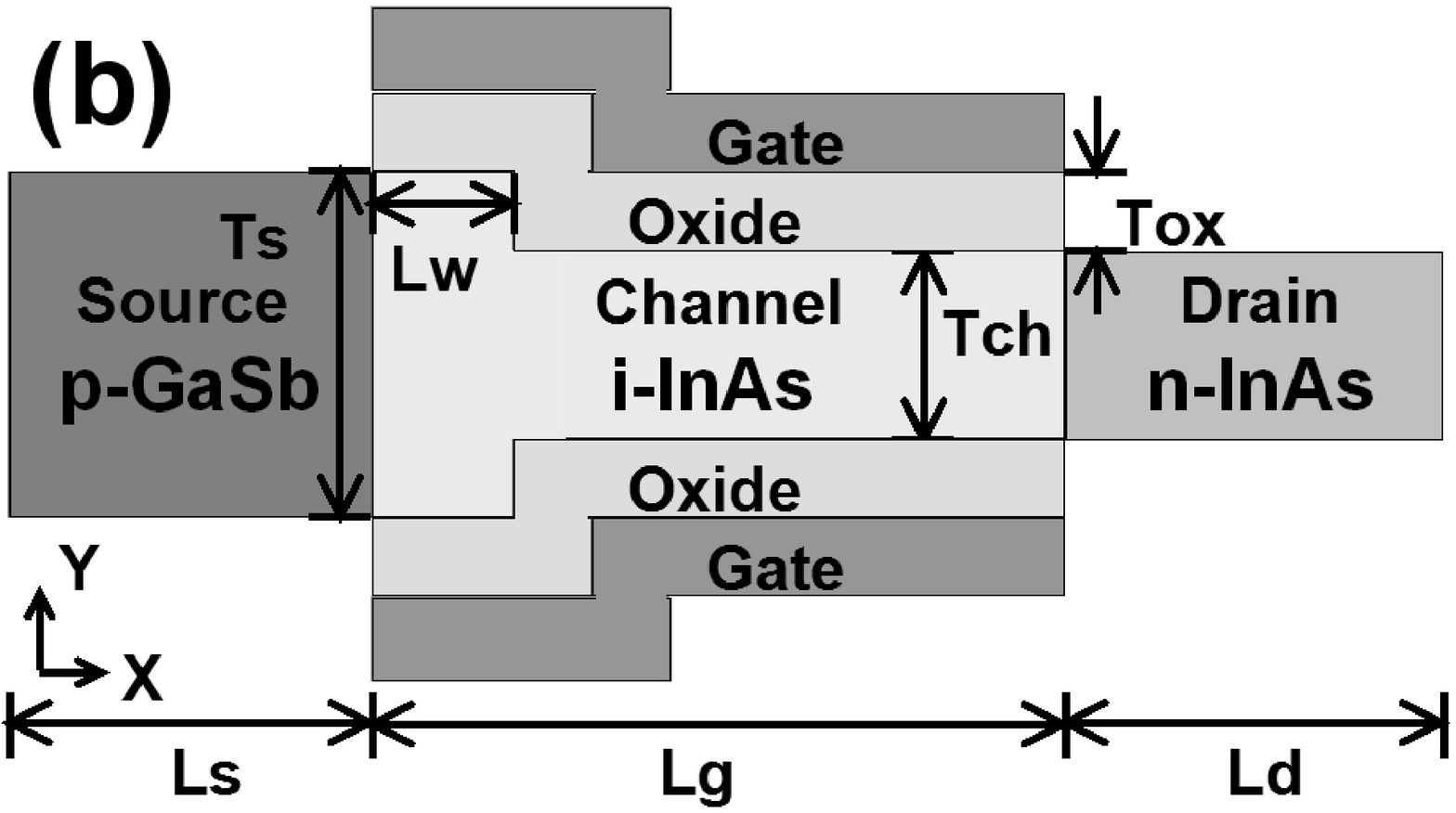}
\caption{\label{fig:geo} GaSb/InAs n-type TFETs with conventional uniform (a) and proposed non-uniform (b) body thickness. The oxide and the gate in (b) are conformal to the channel so that the oxide thickness is equal to that of (a). Here, Tch/Tox is the channel/oxide thickness, $\epsilon_{ox}$ is the oxide dielectric constant, Ls/Lg/Ld is the source/channel/drain length, and Ns/Nd is the source/drain doping density. The extra parameters in (b) are the source thickness Ts and the wide channel length Lw.}
\end{figure}

In this study, we show that by reducing the InAs channel body thickness meanwhile retaining a relatively large body thickness at the source tunnel junction, the leakage can be significantly reduced without compromising the large source tunnel probability. The devices are simulated and optimized by solving Poisson equation and open-boundary Schr\"{o}dinger equation \cite{luisier2006atomistic} self-consistently within NEMO5 tool \cite{steiger2011nemo5}. The Hamiltonian employed is in the atomistic $sp^3d^5s^*$ tight binding (TB) basis including spin-orbit coupling, with the room temperature TB parameters fitted to the band structures as well as the wave functions of density functional theory calculations for better transferability \cite{tan2015tight, Tan2016Online}. Due to the high semiconductor-to-oxide barrier height \cite{robertson2006}, the oxide is treated as an impenetrable potential barrier and only modeled in the Poisson equation. Various scattering mechanisms, such as the electron-phonon scattering \cite{Knoch2010, Luisier2010} and the discrete dopant scattering \cite{Khayer2011}, are not included, which may lead to underestimated SS. Non-ideality effects, such as the interface trap assisted tunneling and the Shockley-Read-Hall generation-recombination processes \cite{Mookerjea2010}, are not considered in this study.

\begin{figure}
\includegraphics[width=8.5cm]{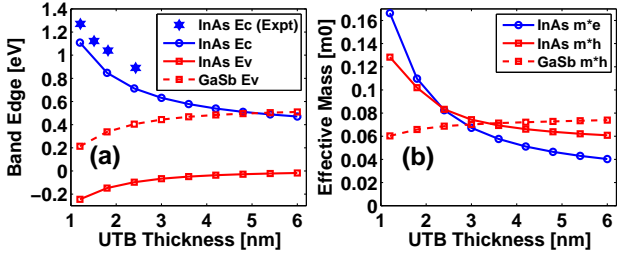}
\caption{\label{fig:edge_mass} (a) Conduction and valence band edges (Ec and Ev), (b) electron and hole effective masses (m*e and m*h), of the GaSb and InAs UTBs, as functions of the UTB thickness. The confinement is in the [001] orientation and the effective masses are in the [100] orientation.}
\end{figure}

\begin{figure}
\includegraphics[width=8.5cm]{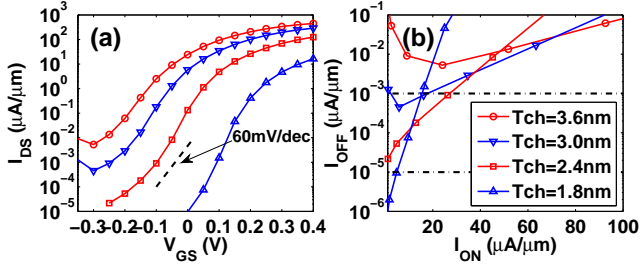}
\caption{\label{fig:IV_uni} (a) $I_{\rm{DS}}$-$V_{\rm{GS}}$ characteristics ($V_{\rm{DS}}$=0.3V) of the uniform device (Fig.~\ref{fig:geo}~(a)) as a function of the body thickness Tch. Confinement/transport is in the [001]/[100] orientation. Lg=15nm, Tox=1.8nm, $\epsilon_{ox}$=9.0, Ns=$-5\times10^{19}/\rm{cm}^{3}$, and Nd=$+2\times10^{19}/\rm{cm}^{3}$. (b) $I_{\rm{ON}}$-$I_{\rm{OFF}}$ curves with $V_{\rm{DD}}=0.3$V.}
\end{figure}

First, we analyze the conventional GaSb/InAs UTB n-type TFETs with uniform body thickness, as illustrated in Fig.~\ref{fig:geo}~(a). The band edges and the effective masses of the ultra-thin-body (UTB) structures for different body thicknesses are plotted in Fig.~\ref{fig:edge_mass}. As can be seen, the InAs UTB band gap (InAs Ec$-$InAs Ev) and the tunnel barrier height (InAs Ec$-$GaSb Ev) both increase as the UTB thickness becomes smaller. The InAs Ec calculated is in agreement with the experiment data \cite{brar1993photoluminescence} with upto 0.2eV mismatch \footnote{The experimental data is shifted by 0.15eV to account for the valence band offset (VBO) between bulk InAs and confined AlSb (with 5nm thickness), assuming the VBO between bulk InAs and bulk AlSb is 0.18eV \cite{vurgaftman2001band} and the confined heavy hole energy of AlSb is 0.03eV. Note that the experiment data were obtained at low temperature.}. The effective masses (electron and hole) of the InAs UTB also increase significantly as the UTB thickness decreases. Therefore, the I-V characteristic of the device is a strong function of the UTB thickness. Indeed, as shown in Fig.~\ref{fig:IV_uni}~(a), a large body thickness (Tch=3.6nm or 3.0nm) leads to a large turn-on current, but also a large SS and a high $I_{\rm{OFF}}$. A small body thickness (Tch=1.8nm) gives rise to a small SS and a low $I_{\rm{OFF}}$, but a small turn-on current. With $I_{\rm{OFF}}=1\times10^{-3}$A/m and $V_{\rm{DD}}=0.3$V, the optimal body thickness for Lg=15nm is around 2.4nm, providing $I_{\rm{ON}}=25$A/m (Fig.~\ref{fig:IV_uni}~(b)). This is too low for any practical logic application.

\begin{figure}
\includegraphics[width=8.5cm]{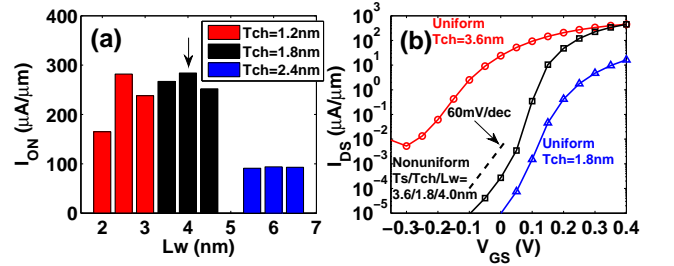}
\caption{\label{fig:IV_uni_non_uni} (a) $I_{\rm{ON}}$ (at $V_{\rm{DD}}$=0.3V and $I_{\rm{OFF}}$=$1\times10^{-3}$A/m) of the nonuniform design (Fig.~\ref{fig:geo}~(b)) for different values of Lw and Tch, with fixed Ts=3.6nm. (b) Full $I_{\rm{DS}}$-$V_{\rm{GS}}$ characteristics of the Ts/Tch/Lw=3.6nm/1.8nm/4.0nm case in (a), in comparison with the uniform 3.6nm and uniform 1.8nm cases. All other device parameters are the same as those in Fig.~\ref{fig:IV_uni}.}
\end{figure}

\begin{figure}
\includegraphics[width=8.5cm]{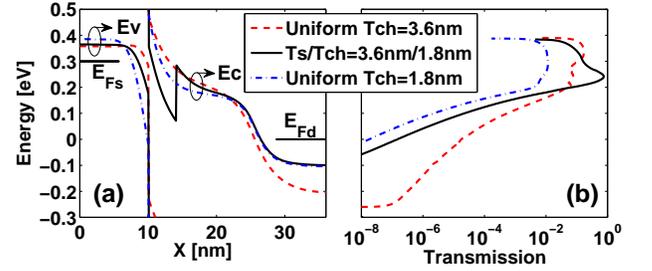}
\caption{\label{fig:band_trans} Comparison of the band diagrams (a) and the transmission functions (b) of the three cases in Fig.~\ref{fig:IV_uni_non_uni}~(b): uniform 3.6nm body thickness, uniform 1.8nm thickness, and non-uniform 3.6nm/1.8nm body thickness.}
\end{figure}

The proposed design that can overcome this dilemma is illustrated in Fig.~\ref{fig:geo}~(b). Compared with Fig.~\ref{fig:geo}~(a), this design has reduced body thickness {\it only} in part of the channel region (and in the drain).
Three parameters need to be optimized, {\it i.e.}, Ts, Tch, and Lw. In this study, we fix Ts to be 3.6nm and then optimize Tch and Lw to maximize $I_{\rm{ON}}$. As shown in Fig.~\ref{fig:IV_uni_non_uni}~(a), there is an optimal Lw for each Tch and the optimal Lw is smaller for smaller Tch, which will be explained in a moment. A tradeoff of Tch is also clearly observed, since a large Tch does not have sufficiently large band gap and effective masses needed for the leakage suppression, while a small Tch would affect the ON-state transmission due to the large reflection at the waveguide discontinuity. With Tch=1.8nm and Lw=4.0nm, we obtain the largest $I_{\rm{ON}}$ (284A/m), which is more than an order of magnitude larger than that of the uniform case (25A/m). The full $I_{\rm{DS}}$-$V_{\rm{GS}}$ curve is further displayed in Fig.~\ref{fig:IV_uni_non_uni}~(b) along with two uniform thickness cases, showing that both steep SS and large turn-on current are simultaneously obtained in the nonuniform case.

The improvements can be better understood from the band diagrams and the transmissions plotted in Fig.~\ref{fig:band_trans}, where the three cases in Fig.~\ref{fig:IV_uni_non_uni}~(b) are compared. Above the channel Ec, the nonuniform 3.6/1.8nm case has larger transmission than the uniform 1.8nm case, giving rise to its larger turn-on current. This is due to its smaller tunneling barrier height and smaller effective masses at the tunnel junction. Its transmission over channel Ec is even larger than the uniform 3.6nm case, benefiting from its larger tunneling electric field and resonance-enhanced tunneling, both resulting from the confined band offset. Below the channel Ec, the nonuniform 3.6/1.8nm case has steeper transmission slope than the uniform 3.6nm case, implying steeper SS. This is partly due to the larger channel band gap and larger channel effective masses, partly due to the better electrostatics at the channel-drain junction, and partly due to the smaller drain Fermi degeneracy (the energy distance between the drain Fermi level and the drain Ec).

\begin{figure}
\includegraphics[width=8.5cm]{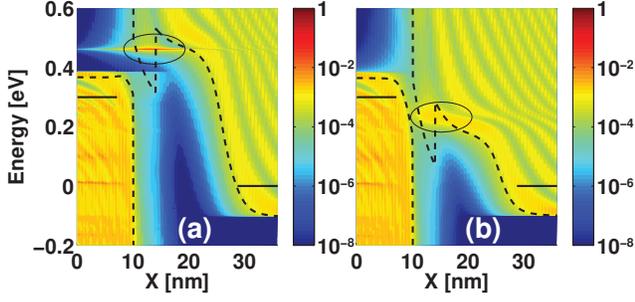}
\caption{\label{fig:ldos} LDOS (in logarithmic scale) of the optimized design in Fig.~\ref{fig:IV_uni_non_uni}, at OFF state (a) and ON state (b). Band diagrams (dashed lines) and contact Fermi levels (solid lines) are superimposed. The quasi-bound states are highlighted (circles).}
\end{figure}

The local denstiy of states (LDOS) is further depicted in Fig.~\ref{fig:ldos}. Similar to the channel heterojunction design \cite{long2016high}, the confined conduction band edges in the channel form a quantum well which creates a quasi-bound state. The energy level of this state needs to be aligned with the channel conduction band edge at the ON state so that it enhances the $I_{\rm{ON}}$. At the OFF state, in order to reduce phonon-assisted tunneling leakage, the energy of this state has to be higher than the valence band edge at the source by at least the optical phonon energy \cite{koswatta2010possibility} ($\hbar\omega_{op}\approx$30meV for bulk InAs and here it is 75meV higher). To satisfy these requirements, the parameter Lw needs to be adjusted for a given Tch, as shown in Fig.~\ref{fig:IV_uni_non_uni}~(a). In fact, a smaller Tch leads to a larger confined band offset and thus the Lw needs to be reduced properly to shift the quasi-bound state upward.

\begin{figure}
\includegraphics[width=8.5cm]{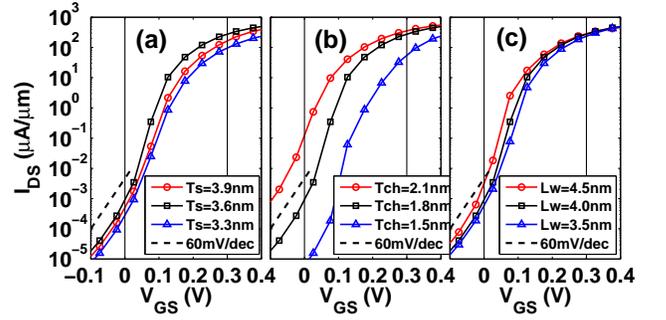}
\caption{\label{fig:variations} Sensitivities of the $I_{\rm{DS}}$-$V_{\rm{GS}}$ curve ($V_{\rm{DS}}$=0.3V) to the variations of Ts (a), Tch (b), and Lw (d), for the optimized design in Fig.~\ref{fig:IV_uni_non_uni}. The amount of Ts and Tch variations is one monolayer (about $\pm0.3$nm) and the Lw variation is $\pm0.5$nm.}
\end{figure}

Fig.~\ref{fig:variations} shows the sensitivities of the I-V curve to the geometry variations. 
We find that the I-V curve can tolerate certain amount of Ts and Lw variations. 
It is however very sensitive to the channel thickness (Tch) variations resulting in unacceptable $I_{\rm{OFF}}$ or $I_{\rm{ON}}$ level. 
This suggests that precise fabrication control of Tch (at atomic level) is necessary.
%that $I_{\rm{OFF}}$ is reduced from $1\times10^{-3}$A/m to $3\times10^{-4}$A/m at Ts=3.3nm and $5\times10^{-4}$A/m at Ts=3.9nm; while $I_{\rm{ON}}$ is also reduced from 284A/m to 100A/m at Ts=3.3nm and 174A/m at Ts=3.9nm. (b) shows that $I_{\rm{OFF}}$ is reduced from $1\times10^{-3}$A/m to $6\times10^{-6}$A/m at Tch=1.5nm and increased to $1\times10^{-1}$A/m at Tch=2.1nm; while $I_{\rm{ON}}$ is reduced from 284A/m to 64A/m at Tch=1.5nm and increased to 360A/m at Tch=2.1nm. (c) shows that $I_{\rm{OFF}}$ is reduced from $1\times10^{-3}$A/m to $6\times10^{-4}$A/m at Lw=3.5nm and increased to $3\times10^{-3}$A/m at Lw=4.5nm; while $I_{\rm{ON}}$ is reduced from 284A/m to 243A/m at Lw=3.5nm and increased to 331A/m at Lw=4.5nm.

\begin{figure}
\includegraphics[width=8.5cm]{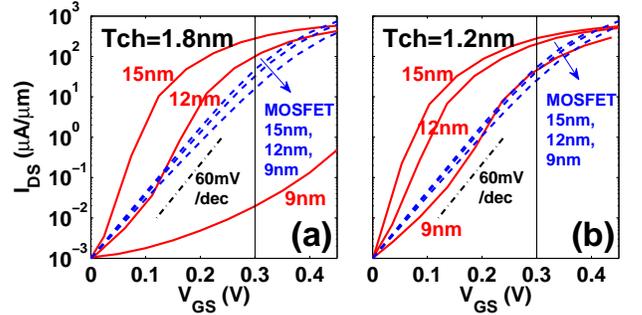}
\caption{\label{fig:scale} $I_{\rm{DS}}$-$V_{\rm{GS}}$ characteristics ($V_{\rm{DS}}$=0.3V) of the proposed TFETs, in comparison with Si MOSFETs, for three channel lengths (Lg=15nm, 12nm, and 9nm). (a) Ts/Tch/Lw=3.6/1.8/4.0nm for TFETs and Tch=1.8nm for MOSFETs, and (b) Ts/Tch/Lw=3.6/1.2/2.5nm for TFETs and Tch=1.2nm for MOSFETs.}
\end{figure}

ITRS 2020 and 2023 technology nodes require channel length Lg to be scaled to about 12nm and 9nm \cite{itrs_website}. At such short channel lengths, the source-to-drain tunneling leakage becomes more prominent. As compared in Fig.~\ref{fig:scale}~(a), when Lg is reduced from 15nm to 12nm, the $I_{\rm{ON}}$ of TFET drops from 284A/m to 106A/m, while the $I_{\rm{ON}}$ of Si MOSFET only drops from 49A/m to 41A/m. When Lg is further scaled to 9nm, the TFET cannot provide a decent ON/OFF ratio and does not possess an advantage over Si MOSFET. To improve the scalability, the channel thickness (Tch) can be reduced from 1.8nm to 1.2nm to enlarge the channel band gap and channel effective masses. As seen in Fig.~\ref{fig:scale}~(b), the SS of TFET degrades less, with $I_{\rm{ON}}$=209A/m (50A/m) obtained at Lg=12nm (9nm), which is still considerably larger than $I_{\rm{ON}}$=47A/m (31A/m) of Si MOSFET.

In summary, a novel transistor design with a nonuniform body thickness is proposed to improve the scalability of GaSb/InAs n-type TFETs. At 15nm channel length, it achieves ballistic $I_{\rm{ON}}$ that is over $10\times$ greater than that is achievable with uniform body thickness. Moreover, it enables the channel to be scaled towards sub-10nm lengths. This design can also be employed to improve InAs/GaSb p-type TFETs.

%\begin{acknowledgments}
This work uses nanoHUB.org computational resources operated by the Network for Computational Nanotechnology funded by the U.S. National Science Foundation under Grant EEC-0228390, Grant EEC-1227110, Grant EEC-0634750, Grant OCI-0438246, Grant OCI-0832623, and Grant OCI-0721680. This material is based upon work supported by the National Science Foundation under Grant 1509394. NEMO5 developments were critically supported by an NSF Peta-Apps award OCI-0749140 and by Intel Corp.
%\end{acknowledgments}

\nocite{*}
\bibliographystyle{apl}
\bibliography{non-uniform}% Produces the bibliography via BibTeX.

\end{document}